\documentclass[aps, prd, twocolumn, amsmath, floats,floatfix, superscriptaddress, nofootinbib]{revtex4}
\usepackage{graphicx}% Include figure files
\usepackage{bm}% bold math
\usepackage{diagbox}
\usepackage{hyperref}% add hypertext capabilities
\usepackage[utf8]{inputenc}
\usepackage{subcaption,booktabs}

\makeatletter\def\Hy@Warning#1{}\makeatother
\hypersetup{
  colorlinks=true,        % false: boxed links; true: colored links
  linkcolor=blue,         % color of internal links
  citecolor=cyan,         % color of links to bibliography
  urlcolor=cyan            % color of external links
}
\usepackage{amsmath,amsfonts,amssymb}
\usepackage{times}
\usepackage[T1]{fontenc}
\usepackage{float}
\usepackage[normalem]{ulem}
\usepackage{multirow}
\usepackage{soul}
\usepackage[justification=raggedright,singlelinecheck=false]{caption}

\newcommand{\pr}[1]{\ensuremath{\left[#1\right]}}
\newcommand{\pc}[1]{\ensuremath{\left(#1\right)}}

\DeclareMathOperator*{\argmin}{arg\,min} % thin space, limits underneath in displays

\newcommand{\be}{\begin{equation}}
\newcommand{\ee}{\end{equation}}
\newcommand{\bea}{\begin{eqnarray}}
\newcommand{\eea}{\end{eqnarray}}

\makeatletter
\newcommand*{\rom}[1]{\expandafter\@slowromancap\romannumeral #1@}
\makeatother

\makeatletter
\newcommand{\thickhline}{%
    \noalign {\ifnum 0=`}\fi \hrule height 1pt
    \futurelet \reserved@a \@xhline
}
\newcolumntype{"}{@{\hskip\tabcolsep\vrule width 1pt\hskip\tabcolsep}}
\makeatother
\begin{document}
\title{
Detecting Hyperons in neutron stars -- a machine learning approach}

 \author{Valéria Carvalho}
 \email{val.mar.dinis@uc.pt}
 \affiliation{CFisUC, 
 	Department of Physics, University of Coimbra, P-3004 - 516  Coimbra, Portugal}

 \author{Márcio Ferreira}
 \email{marcio.ferreira@uc.pt}
 \affiliation{CFisUC, 
 	Department of Physics, University of Coimbra, P-3004 - 516  Coimbra, Portugal}
	
 \author{Constança Providência}
 \email{cp@uc.pt}
 \affiliation{CFisUC, 
 	Department of Physics, University of Coimbra, P-3004 - 516  Coimbra, Portugal}

\date{\today}

\begin{abstract} 
We present a neural network classification model for detecting the presence of hyperonic degrees of freedom in neutron stars. The models take radii and/or tidal deformabilities as input and give the probability for the presence of hyperons in the neutron star composition.  Different numbers of observations and different levels of uncertainty in the neutron star properties are tested.  The models have been trained on a dataset of well-calibrated microscopic equations of state of neutron star matter based on a relativistic mean-field  formalism. Real data and data generated from a different description of hyperonic matter are used to test the performance of the models.

\end{abstract}

\maketitle

\section{Introduction}

Neutron stars (NSs) are among the densest objects in the Universe, yet their internal properties and composition remain open questions. The equation of state (EOS) of NS matter, dense and asymmetric nuclear matter, is the main research focus in NS physics. In particular, observations of massive NSs provide constraints on the EoS at intermediate and high baryonic densities. In addition, the advent of multi-messenger astrophysics, which combines different sources of information about NS such as gravitational waves (GWs), photons, and neutrinos, has improved our knowledge of NS physics.\\

At moderate and high baryonic densities, constraints on the EOS of NS matter are primarily provided by observations of massive NSs. Key measurements include $1.908 \pm 0.016 M_{\odot}$ for PSR J1614-2230 \cite{Demorest2010,Fonseca2016,Arzoumanian2017}, $2.01 \pm 0.04 M_{\odot}$ for PSR J0348-0432 \cite{Antoniadis2013}, $2.08 \pm 0.07 M_{\odot}$ for PSR J0740+6620 \cite{Fonseca:2021wxt}, and $2.13 \pm 0.04 M_{\odot}$ for PSR J1810+1714 \cite{Romani:2021xmb}. The rise of multi-messenger astrophysics has further enriched our understanding of NS physics by combining information from various sources, including GWs, photons, and neutrinos. The detection of compact binary coalescence events, such as GW170817 \cite{Abbott:2018wiz} and GW190425 \cite{Abbott:2020khf} by the LIGO/Virgo collaboration, has placed additional constraints on the EOS. Moreover, recent mass and radius inferences from the NICER (NS Interior Composition ExploreR) experiment, including PSR J0030+045 \cite{Riley_2019,Miller19}, and  PSR J0740+6620's radius \cite{Riley2021,Miller2021,Raaijmakers2021}
 have significantly narrowed the range of possible NS matter scenarios. 
 Additionally, observations of PSR J0427-4715 \cite{Choudhury:2024xbk,reardon2024neutron}, the brightest and closest known millisecond pulsar, provide further constraints on both mass and radius, complementing the previous measurements.
Future experiments, with enhanced precision, are expected to further refine our understanding of the EOS and NS properties. Missions such as the enhanced X-ray Timing and Polarimetry mission (eXTP) \cite{eXTP,eXTP:2018anb}, the Spectroscopic Time-Resolving Observatory for Broadband Energy (STROBE-X) \cite{STROBE-X}, and the Square Kilometer Array (SKA) telescope \citep{SKA} will be crucial in constraining the different theoretical scenarios for NS matter.\\

From the theoretical side, the low-density region of the NS EOS is constrained by chiral effective field theory (cEFT) \cite{Hebeler2013,Drischler:2017wtt,Huth:2021bsp}, while perturbative quantum chromodynamics (pQCD)  becomes reliable at high densities \cite{Kurkela2009} (see \cite{ghiglieri2020perturbative} for a review). Inference of the NS EOS, given a set of  theoretical, experimental and observational constraints, is frequently carried out using Bayesian inference frameworks \cite{Landry:2018prl,Traversi_2020,Raaijmakers2021,Essick:2021ezp,Huth:2021bsp,Malik:2022zol,Mondal:2022cva,Gorda:2022jvk,Sun:2022yor,Huang:2023grj,Malik:2023mnx}. Several studies take an agnostic modeling of the EOS, e.g., piecewise polytropes, the speed of sound formalism, spectral approaches, Gaussian processes \cite{Kurkela:2014vha,Annala:2017llu,Landry:2018prl,Tews:2018iwm,Annala2019,Essick2019,Somasundaram:2021clp,Altiparmak:2022bke,Gorda:2022jvk}. Although these methods provide valuable insights into the EOS, they do not offer information about the underlying particle composition of NS matter.\\

NSs are composed of five distinct layers: the atmosphere, outer crust, inner crust, outer core, and inner core. However, the exact composition of the core remains a subject of ongoing investigation. Several hypotheses have been proposed, suggesting that the core could consist of nucleonic matter, quark matter, hyperons, or even dark matter. One of the most discussed possibilities is the presence of hyperons in the inner core, see for example \cite{Bednarek:2011gd,Weissenborn:2011kb,Oertel:2014qza,Chatterjee:2015pua,tolos2020strangeness,malik2022bayesian,providencia2023neutron}. Due to the extreme densities at the center of an NS and the rapid increase in the chemical potential of the nucleon with increasing density, the formation of strange hadrons, such as hyperons, could become energetically favorable. While the appearance of hyperons seems almost inevitable under such conditions, their presence would soften the EOS, which could lead to a maximum NS mass that contradicts the observed $2M_{\odot}$ NSs \cite{Vidana:2010ip,Schulze:2011zza,Lonardoni:2014bwa,Chatterjee:2015pua}. This discrepancy, known as the hyperon puzzle, remains unsolved.\\

The use of machine learning frameworks, and specially feed-forward neural networks (NNs) based methods, as inference tools in high-energy physics has seen a growing surge of interest across various disciplines \cite{Zhou:2023pti}.
In NS physics, 
NNs have been extensively applied to study the EoS of NSs in several works \cite{Ferreira:2022nwh,ferreira2021unveiling,Fujimoto_2021,Soma:2022qnv,morawski2020neural,PhysRevC.106.065802,han2021bayesian}. In addition, Bayesian NNs, which are models capable of uncertainty quantification, have been employed to map NS observations to critical quantities such as the speed of sound squared and the proton fraction within NS cores \cite{Carvalho:2023ele}, as well as to investigate the properties of nuclear matter \cite{PhysRevD.109.123038}. 
Recently, an upgraded method for obtaining the probability distribution of NS $M(R)$ relationships was introduced in \cite{Fujimoto:2024cyv}, improving upon earlier works \cite{fujimoto2021extensive,Fujimoto_2020,Fujimoto_2018}. This new approach uses Monte Carlo integration rather than Gaussian input noise to refine predictions, while also incorporating new observational data from NICER and the GW170817 event.
Generative modelling has also begun to emerge as a promising alternative for decoding NS characteristics. For instance, the use of normalizing flows to detect phase transitions in NSs through anomaly detection \cite{Carvalho:2024zyb}, or the use of conditional variational autoencoders as EOS sampler (conditional on an observational set) to gain insights into NS properties \cite{Ferreira:2024rnf}. 
A framework aimed to infer the EOS 
directly from telescope observations based on normalizing flows model coupled with Hamiltonian Monte Carlo methods was introduced in \cite{Brandes:2024vhw}.
Studies focused on GW also play a significant role in constraining the EoS of NS. Machine learning can be used to analyze GW signals from binary NS mergers. 
For instance, \cite{Goncalves:2022smd} explored the use of the Audio Spectrogram Transformer model, a type of NN architecture inspired by how we process sound, to classify the EoS solely based on GW signals. \\

This work investigates the potential identification of the presence of hyperons within NSs by employing NN classification models to analyze key NS observables. Specifically, we explore three types of input data: mass-radius, mass-tidal deformability, and a combination of both, while also varying the level of input noise. The training dataset consists of relativistic mean-field (RMF) models, which are constrained by minimal conditions such as low-density properties and pure neutron matter, in addition to describing two solar-mass stars using Bayesian inference.
To evaluate the models' performance, we first tested the trained models using two theoretically based datasets: one derived from the same dataset used for training, and another generated with a different formalism but also constructed within the RMF framework. Following this, we assessed the models' predictive capabilities for mass-radius observables, utilizing observational data from our previous study \cite{PhysRevD.109.123038}. The observational data were carefully adjusted to match the respective training regions, ensuring consistency between the theoretical and real-world test cases.
The paper is organized as follows: Section \ref{dataset} introduces the selected family of nuclear models and explains the generation of mock observational datasets. Section \ref{NN} provides a brief overview of NNs.  The results are presented and discussed in Section \ref{results}, followed by concluding remarks in Section \ref{conclusion}.

\section{Dataset \label{dataset}}
To achieve our goal of identifying the potential existence of hyperons within NSs, we first need to construct a comprehensive dataset, which will serve as the foundation for developing our machine learning models. This section outlines the process of creating the different datasets used throughout this study.

\subsection{Theoretical concepts of dataset}
In the present study we adopt the description of baryonic matter within a RMF approach \cite{book.Glendenning1996}, a framework which allows to include hyperonic degrees of freedom in a straightforward way. The EOS is obtained from a Lagrangian density written in terms of the Dirac spinors ${\Psi}$ that describe nucleons, and hyperons in the case of hyperonic matter, as well as leptons to describe $\beta$-equilibrium and electrically neutral matter.  The baryonic interaction is described in terms of an exchange of mesons: a scalar isoscalar $\sigma$ field, a vector isoscalar $\omega$ field,  a vector isovector $\varrho$ field, and in the case of hyperonic matter a vector isoscalar $\phi$ field with hidden strangeness. The meson  masses are  $m_i,\, i=\sigma, \, \omega,\, \varrho,\, \phi$.  The parameters $g_i$, $i=\sigma, \, \omega,\, \varrho,\, \phi$ designate the couplings of the mesons to the baryons.
The dataset used in the present study derives from the Lagrangian density 
\begin{equation}
\begin{aligned}
\mathcal{L}=& \bar{\Psi}\Big[\gamma^{\mu}\left(i \partial_{\mu}-g_{\omega} A_{\mu}^{(\omega)}-%\frac{1}{2}
g_{\varrho} {\boldsymbol{t}} \cdot \boldsymbol{A}_{\mu}^{(\varrho)}\right) %\\&
-\left(m-g_{\sigma} \phi\right)\Big] \Psi  \\&
+ \frac{1}{2}\left(\partial_{\mu} \phi \partial^{\mu} \phi-m_{\sigma}^{2} \phi^{2} \right) \\
&-\frac{1}{4} F_{\mu \nu}^{(\omega)} F^{(\omega) \mu \nu} 
+\frac{1}{2}m_{\omega}^{2} A_{\mu}^{(\omega)} A^{(\omega) \mu} \\&
-\frac{1}{4} \boldsymbol{F}_{\mu \nu}^{(\varrho)} \cdot \boldsymbol{F}^{(\varrho) \mu \nu} 
+ \frac{1}{2} m_{\varrho}^{2} \boldsymbol{A}_{\mu}^{(\varrho)} \cdot \boldsymbol{A}^{(\varrho) \mu}+ \mathcal{L}_{NL}.
\end{aligned}
\label{lagrangian}
\end{equation}
The last term $\mathcal{L}_{NL}$ includes all the non-linear (NL) mesonic terms and is absent in frameworks with density dependent couplings $g_i$. Our dataset is generated considering this term in the Lagrangian density together with constant $g_i$ meson-baryon couplings, this approach and the corresponding EOS  are referred to as NL.  An alternative would be to consider density dependent meson couplings \cite{Typel1999}. In Sec. \ref{DDB} a set of EOSs obtained within this last formalism in \cite{Malik:2022zol} will also be introduced and referred to as DDB EOSs. In the above expression  $\gamma^\mu $ and $\boldsymbol{t}$ designate, respectively, the Dirac matrices and the isospin operator. The  vector meson field strength tensors are given by  $F^{(\omega, \varrho)\mu \nu} = \partial^ \mu A^{(\omega, \varrho)\nu} -\partial^ \nu A^{(\omega, \varrho) \mu}$.  

For the  term $\mathcal{L}_{NL}$  we take
\begin{eqnarray}
  \mathcal{L}_{NL}=&-\frac{1}{3} b g_\sigma^3 (\sigma)^{3}-\frac{1}{4} c g_\sigma^4 (\sigma)^{4}+\frac{\xi}{4!}(g_{\omega}^2\omega_{\mu}\omega^{\mu})^{2} \nonumber \\ &+\Lambda_{\omega}g_{\varrho}^{2}\boldsymbol{\varrho}_{\mu} \cdot \boldsymbol{\varrho}^{\mu} g_{\omega}^2\omega_{\mu}\omega^{\mu}.
  \label{lagrangian_nl}
\end{eqnarray}
 The couplings in front of each term,  $b,\, c,$ $\xi$, $\Lambda_\omega$ are fixed by a given set of values together with the meson-nucleon couplings $g_i$  imposing the constraints defined in Table \ref{tab1} within a Bayesian inference calculations, see \cite{Malik:2023mnx} for more details. This Lagrangian density is the starting point of many RMF models \cite{Horowitz:2000xj,Todd-Rutel2005,Providencia:2013dsa,Chen:2014sca,Bao:2015cfa,Tolos:2017lgv,Fattoyev:2020cws}.
 The couplings in front of each term,  $b,\, c,$ $\xi$, $\Lambda_\omega$ are fixed together with the meson-nucleon couplings $g_i$  imposing the constraints defined in Table \ref{tab1} within a Bayesian inference calculations, see \cite{Malik:2023mnx} for more details. This Lagrangian density is the starting point of many RMF models \cite{Horowitz:2000xj,Todd-Rutel2005,Providencia:2013dsa,Chen:2014sca,Bao:2015cfa,Tolos:2017lgv,Fattoyev:2020cws}.
 
The introduction of hyperons requires that the first term of the Lagrangian density, eq. (\ref{lagrangian}), also sums over the hyperons.  The couplings of the hyperons to the mesons were constrained by the properties of the hypernuclei and implemented as described in \cite{Malik:2022jqc}: for the vector isoscalar mesons, the SU(6) quark model values were chosen. For the vector isovector, the interaction is defined by the isospin projection multiplied by the $\varrho$ meson-nucleon coupling. The possible range of values for the coupling to the sigma meson has been obtained from a rather large number of RMF models by fitting some hypernuclei properties (see \cite{Fortin:2017cvt,Providencia:2018ywl,Fortin:2020qin}).   Note that only the most probable hyperons, $\Lambda$ and $\Xi^-$, were considered. The experimental data seem to indicate that the $\Sigma$-nucleon interaction is repulsive, which prevents the onset of the $\Sigma^-$ hyperon inside NS.

After the detection of the two solar mass pulsar PSR J1614-2230 \cite{Demorest2010},  it was argued that baryonic matter with hyperons would not support two solar mass NS, see discussion in \cite{Demorest2010,Vidana:2010ip}. However, it was later shown that two solar mass stars could also include hyperons not only within a RMF description considering the interaction constrained by hypernuclei properties \cite{Bednarek:2011gd,Weissenborn:2011kb,Providencia:2013dsa,Oertel:2014qza}  but also within a microscopic non-relativistic  description solving the Schrödinger equation by means of an auxiliary field diffusion Monte Carlo algorithm and including two-body and three body forces involving the $\Lambda$ hyperon \cite{Lonardoni:2013gta}.  For a more complete overview of these subject, we refer the reader to the reviews \cite{Chatterjee:2015pua,tolos2020strangeness,Burgio:2021vgk}. In our study, we use datasets generated within the RMF description which allows for two solar mass stars with hyperons.

The data set, which is publicly available\footnote{\url{https://zenodo.org/records/7854112}}, contains 17810 EoS of pure nucleonic matter and 18728 EoS of hadronic matter containing both nucleons and hyperons. 

\begin{figure*}[!hbt]
    \centering
\includegraphics[width=0.33\linewidth]{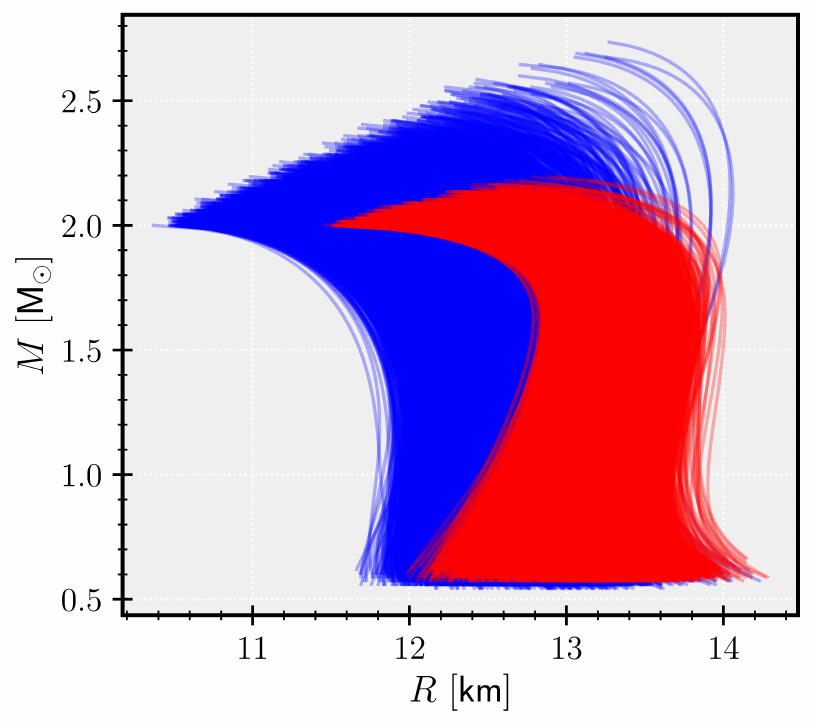}
\includegraphics[width=0.33
    \linewidth]{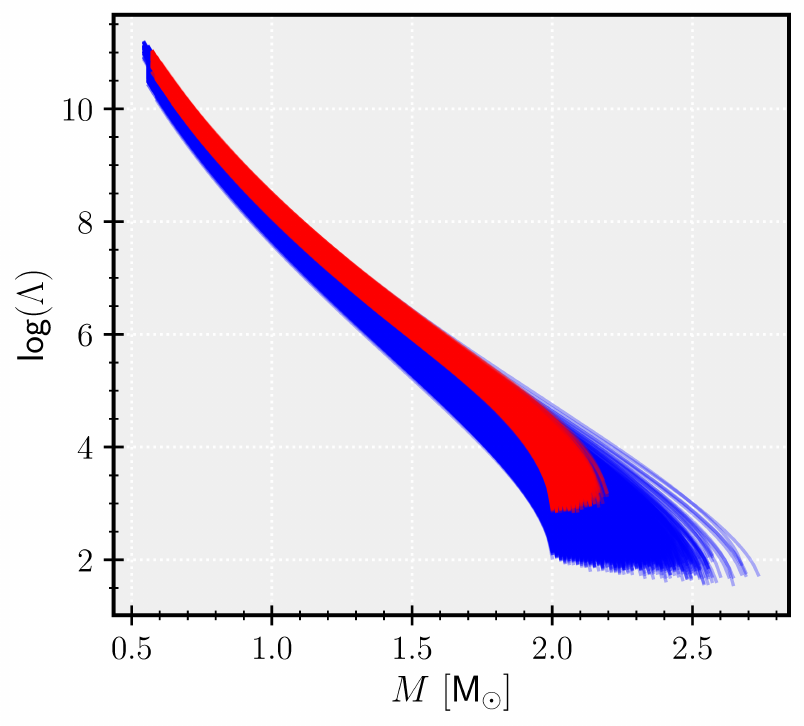}
    \includegraphics[width=0.33
    \linewidth]{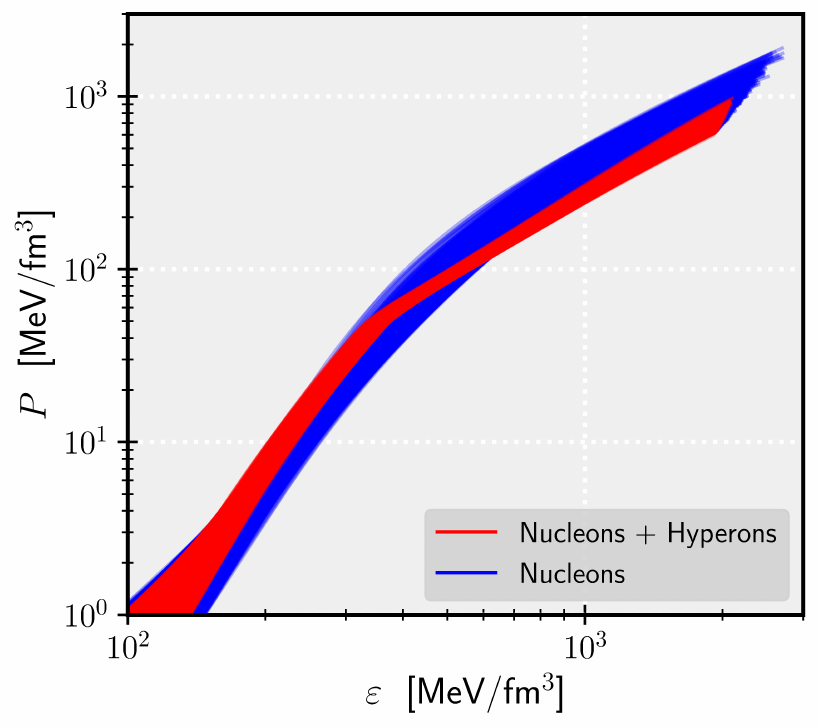}
    \caption{Mass-radius, left plot, mass-log(tidal deformability), center plot, { \color{blue}and pressure-energy density, right plot, } curves for nucleonic matter (17810 curves, shown in blue) and hyperonic matter (18728 curves, shown in red).}
    \label{fig:M_r}
\end{figure*}

{The  properties mass and radius of  spherically symmetric and static relativistic stars in hydrostatic equilibrium
are  obtained by solving the Tolman-Oppenheimer-Volkov (TOV) equations  \cite{PhysRev.55.364,PhysRev.55.374} 
\begin{multline}\label{eq:tov1}
\frac{dP(r)}{dr}=\frac{ \varepsilon(r)m(r)}{r^2}\pr{1+\frac{P(r)}{\varepsilon(r)}}\\ \pr{1+\frac{4\pi r^3 P(r)}{m(r)}}\pr{1-\frac{2m(r)}{r}}^{-1} \;,
\end{multline}
\begin{align}\label{eq:tov2}
\frac{dm(r)}{dr}=& 4\pi r^2\varepsilon(r)\;.
\end{align} 
where $r$ is the radial distance from the NS center, $p(r)$ and $\varepsilon(r)$ are the pressure and energy density inside the NS at $r$, and $m(r)$ is the mass enclosed within the sphere of radius $r$. 
The tidal deformability, $\Lambda$, is expressed in terms of the dimensionless tidal Love number, $k_2$
 , as follows

\begin{equation}\label{eq:tidal}
    \Lambda = \frac{2}{3}k_2(C)^{-5} \;,
\end{equation}
\noindent where $C=M/R$ is the compactness of the star.
The second order Love number is then defined as

\begin{align}\label{eq:love}
    k_2=& \frac{8C^5}{5}\pc{1-2C}^2 \pr{2 + 2C\pc{y_R-1} - y_R}\times  \nonumber\\ 
    &\{ 2C\pr{6-3y_R + 3C\pc{5y_R-8}} \nonumber\\
    &+ 4 C^3 \pr{ 13 - 11y_R + C\pc{3y_R-2} + 2C^2\pc{1+y_R}}\nonumber \\ 
    & +3\pc{1-2C}^2 \pr{2-y_R+2C\pc{y_R-1}} \ln \pc{1-2C}\}^{-1} \;, 
\end{align}

\noindent the boundary value of $y(r)$ at the surface, denoted as $y_R\equiv y(r=R)$ arises from the following first-order differential equation (\cite{Hinderer2008})
\begin{align}
    \frac{dy(r)}{dr}=- \frac{y(r)^2}{r}- \frac{y(r)}{r}F(r) - rQ(r)\; 
\end{align}
\noindent where $F(r)$ and $Q(r)$ being functions of $\varepsilon(r)$, $P(r)$, and $m(r)$, are represented as
\begin{align}
    F(r)&=\pc{1-4\pi r^2\pr{\varepsilon(r)-P(r)}}\pr{1-\frac{2m(r)}{r}}^{-1} \;, \label{eq:td1} \\ 
    Q(r)&=4 \pi \pr{5 \varepsilon(r) + 9P(r) + \frac{\varepsilon(r)+P(r)}{v_s^2(r)} - \frac{6}{r^2}} \nonumber \\
    &\pr{1-\frac{2m(r)}{r}}^{-1} - \frac{4m^2(r)}{r^4}\Biggl[ 1+
    \frac{4\pi r^3 P(r)}{m(r)}\Biggr]^2 \nonumber\\ & \pr{1-\frac{2m(r)}{r}}^{-2}\; .\label{eq:td2}
\end{align}
These equations can be solved numerically simultaneously with the TOV Eqs. \ref{eq:tov1} and \ref{eq:tov2}, also taking into account $y(r=0)=2$, as shown in \cite{Krastev_2022}, along with the conditions for the center of the star. By doing so, the value of $y_R$ is determined and subsequently substituted into the Love number, Eq.\ref{eq:love}, to obtain the tidal deformability  ($\Lambda$) using Eq.\ref{eq:tidal}. 
% In Fig. \ref{fig:M_r} we are able to see in the left the mass-radius curves, in the center the mass-tidal deformability, and in the right plot the $\beta$-equilibrium NS matter EOS with and without hyperons is shown with the underlying EOS which, after integration of the TOV equations, yields the mass-radius distribution shown in Fig. \ref{fig:M_r}. 

In Fig.\ref{fig:M_r}, in the right plot, the $\beta$-equilibrium NS matter EOS with and without hyperons is shown. This underlying EOS, when integrated using the TOV equations alongside  Eqs. \ref{eq:td1} and \ref{eq:td2}, results in the mass-radius distribution present in the left plot and the corresponding mass-tidal deformability curves shown in the center plot.

To solve the TOV equations, a complete EOS with crust and core must be constructed. For the outer crust, the BPS EOS was considered. The inner crust was described by a polytropic function interpolating between the outer crust and the neutron star core at a density of 0.04fm$^{-3}$, see \cite{Malik:2023mnx} for details. This prescription has been shown to give rise to radius uncertainties for low/medium mass stars below 200m, and a negligible effect for massive stars.

The inclusion of hyperons leads to a stiffening of the core EOS at low/intermediate densities, followed by a softening once hyperon onset occurs. The stiffening at low densities is essential for the model to reach $2 M_\odot$ NSs. The behavior of the EOS at low and high densities is directly reflected in the mass-radius and mass-tidal deformability curves, in particular: i) low/medium-mass hyperonic stars can have larger radii and tidal deformabilities than nucleonic stars; ii) the inclusion of hyperons leads to lower maximum masses. 
}

The Bayesian framework used to generate the two sets of EOSs imposes minimal constraints on various nuclear saturation properties. This approach ensures the reproduction of $2M_\odot$ NSs and maintains consistency with low-density pure neutron matter as predicted by N$^3$LO calculations in chiral effective field theory (see table \ref{tab1}). 

\begin{table}[!hbt]
 \caption{The different constraints imposed via Bayesian inference: the binding energy per nucleon $\epsilon_0$, incompressibility $K_0$, and symmetry energy $J_{\rm sym}$ at the nuclear saturation density $n_0$ (with 1$\sigma$ uncertainties). The pressure of pure neutron matter (PNM) is considered at densities of 0.08, 0.12, and 0.16 fm$^{-3}$, obtained from a $\chi$EFT calculation \cite{Hebeler2013}.}
  \label{tab1}
      \centering
 \setlength{\tabcolsep}{5.5pt}
      \renewcommand{\arraystretch}{1.1}
\begin{tabular}{cccc}
\hline 
\hline 
\multicolumn{4}{c}{Constraints}                                                        \\ 
\multicolumn{2}{c}{Quantity}                     & Value/Band  & Ref   \vspace{0.1cm}\\ \hline
\multirow{3}{*}{\shortstack{NMP \hspace{0.cm} }}  {[}fm$^{-3}${]}  & $n_0$ & $0.153\pm0.005$ & \cite{Typel1999}    \\
                                                & $\epsilon_0$ & $-16.1\pm0.2$ & \cite{Dutra:2014qga}   \\
                \hspace{1cm}   {[}MeV{]}        & $K_0$           & $230\pm40$   & \cite{Shlomo2006,Todd-Rutel2005}    \\
                              & $J_{\rm sym, 0}$           & $32.5\pm1.8$  & \cite{Essick:2021ezp}   \\
                              
                               &                 &                &                                                   \\
  \shortstack{PNM \\ {[}MeV fm$^{-3}${]}}                  & $P(\rho)$       & $2\times$ N$^{3}$LO    & \cite{Hebeler2013}   \\
  &$dP/d\rho$&$>0$&\\
%                               &                 &                &                           &                           \\
\shortstack{NS mass \\ {[}$M_\odot${]}}        & $M_{\rm max}$   & $>2.0$     &  \cite{Fonseca:2021wxt}      \\ 

\hline 

\end{tabular}
\end{table}

\subsection{Mock datasets generation}

The goal of this work is to use simulated NS observations, represented by the set $\bm{X}$, as inputs to a machine learning model aimed at classifying whether or not an NS contains hyperons, represented by the set $\bm{Y}$. The input set $\bm{X}$ consists of $\mathcal{D}$ rows of vectors, denoted by $\bm{x}$, while the output set $\bm{Y}$ consists of $\mathcal{D}$ rows of scalars, denoted by $y$. The value of $\mathcal{D}$ defines the size of the data set. To clarify, the output set $\bm{Y}$ is expressed as $\bm{Y}= \{y^{(i)}\}_{i=1}^\mathcal{D}$, and the input set $\bm{X}$ is expressed as $\bm{X}= \{\bm{x}^{(i)}\}_{i=1}^\mathcal{D}$, i.e, the EOSs in the data set are characterized by the tuples $(\bm{x}^{(i)},y^{(i)})$. The output scalar $y$ is binary, where 0 indicates the presence of hyperons and 1 indicates their absence. The input vectors $\bm{x}$ can take three different forms, depending on the data type ($\bm{R}$, $\bm{\Lambda}$, or $\bm{R\Lambda}$) or the number of NS considered ($Q=5,10$ or 15):
\begin{enumerate}
    \item The $\bm{R}$ datasets contain simulated $R(M)$ data. The observations along the $R(M)$ curve are denoted by pairs $(M_q^{R}, R_q)$, and the input vectors take the form $\bm{x}^{R}=\{(M_1^{R},R_1),(M_2^{R},R_2),...,(M_Q^{R},R_Q)\}$, with dimension of 10, 20, and 30, when considering 5, 10, 15 NS, respectively. 
    
    \item The $\bm{\Lambda}$ datasets contain simulated $\Lambda(M)$ data. The observations along the $\Lambda(M)$ curve are denoted by pairs $(M_q^{\Lambda}, \Lambda_q)$, and the input vectors take the form $\bm{x}^{\Lambda}=\{(M_1^{\Lambda},\Lambda_1),(M_2^{\Lambda},\Lambda_2),...,(M_Q^{\Lambda},\Lambda_Q)\}$, with dimension of 10, 20, and 30, when considering 5, 10, 15 NS, respectively. 
    \item The $\bm{R\Lambda}$ datasets contain both simulated $R(M)$  and $\Lambda(M)$ data. The observations along the $R(M)$  and $\Lambda(M)$ curves are denoted by pairs $(M_q^{R}, R_q)$ and $(M_q^{\Lambda}, \Lambda_q)$, and the input vectors take the form $\bm{x}^{R\Lambda}=\{(M_1^{R},R_1,M_1^{\Lambda},\Lambda_1),(M_2^{R},R_2,M_2^{\Lambda},\Lambda_2),...,\\(M_Q^{R},R_Q,M_Q^{\Lambda},\Lambda_Q)\}$, with dimension of 20, 40, and 60, when considering 5, 10, 15 NS, respectively.
\end{enumerate}
The use of three types of datasets is aimed to investigate how much information is available in NSs observations, GW observations, and the combination of both when inferring the possible presence of hyperons in NS.
To ensure a balanced representation of each composition in the dataset, the larger number of hyperonic EoS was adjusted by selecting 17810 EOS from both nucleonic and hyperonic compositions, resulting in a total of 35620 EOS, meaning a  $\cal D$=35620. The dataset was then randomly divided into a training set containing 90\% of the data and a test set comprising the remaining 10\%. \\

The statistical procedure for generating the mock data from the $M(R)$ curves with distinct input noises follows the following steps, as detailed in \cite{PhysRevD.109.123038}. For a given EoS, we first randomly sample $Q$ NS mass values, $M_q^{0} \sim \mathcal{U}{[1,M_{\text{max}}]}$, {where $M_{\text{max}}$ is the maximum mass corresponding to the respective TOV curve,} and determine the radius from the TOV solution, $(M_q^{0},R(M_q^{0}))$, for $q=1,...,Q$. Then, the actual noisy observation values $(M_q^R,R_q)$ are obtained by sampling from Gaussian distributions centered at the TOV solution:  $M_q^R \sim \mathcal{N}(M_q^{0},\sigma_{q, M^R}^2 )$ and $R_q \sim \mathcal{N}(R(M_q^{0}),\sigma_{q,R}^2 )$, where $\sigma_{q,R} \sim \mathcal{U}{[0,\sigma_R]}$ and $\sigma_{q,M^R} \sim \mathcal{U}{[0,\sigma_{ M^R }]}$ for $q=1,...,Q$. The values of $\sigma_R$ and $\sigma_{ M^R }$ are shown in Table \ref{tab:Sets}. These $Q$ pairs $(M_q^R, R_q)$ collectively characterize the $M(R)$ diagram of a given EoS. Hereafter, for a given EOS, we denote
by {\it single observation} the corresponding input vector $\bm{x} = [M_1^R, \cdots, M_Q^R, R_1, \cdots, R_Q]$. 
To incorporate the tidal deformability, we take a similar statistical procedure: we first sample $M_q^{\Lambda} \sim \mathcal{U}{[1,M_{\text{max}}]}$ and then $\Lambda_q \sim  \mathcal{N} ( \textbf{$\Lambda$} ( M_q^{\Lambda}  ),\sigma_{\Lambda}^2( M_q^{\Lambda}  ))$, where $\textbf{$\Lambda$} ( M_q^{\Lambda}  )$ represents the tidal deformability of the star, and the values $\sigma_{\Lambda}^2( M_q^{\Lambda} )$ can be found in  Table \ref{tab:Sets}.  The value of $\sigma_{\Lambda}( M_q^{\Lambda} )$ is defined as  $\sigma_{\Lambda}( M_q^{\Lambda} ) = \text{constant}\times \hat{\sigma}(M_q^{\Lambda})$, where $\hat{\sigma}(M_q^{\Lambda})$ is computed from the training dataset as the standard deviation of $\Lambda(M)$, we use this value given the broad range of values of the tidal deformability.
For datasets containing tidal deformability, a single observation consist of an input structure as $\bm{x} = [M^\Lambda_1, \cdots, M^\Lambda_Q, \Lambda_1, \cdots, \Lambda_Q]$, while when both mass-radius and tidal deformability are considered becomes $\bm{x} = [M_1^R, \cdots, M_Q^R, R_1, \cdots, R_Q,M^\Lambda_1, \cdots, M^\Lambda_Q, \Lambda_1, \cdots, \Lambda_Q]$. During training, the logarithm of tidal deformability is used due to its vastly different scale compared to mass and radius.\\ 

For each EOS, we replicate the above procedure $n_{\text{s}}$ times, i.e.,  the input vector $\bm{x}$ is resampled $n_{\text{s}}$ times. This approach expands the dataset to $\mathbb{D}=n_s \times \mathcal{D}$. For training, each EoS was simulated with 20 mock observations ($n_s=20$), whereas for testing $n_s=1$ was used. This setup mimics a real-world scenario where typically only a single mock observation of the "true" EoS is available.\\

Independent datasets were generated with distinct input noise levels for each input size. Therefore, for each dataset type ($\bm{R}$, $\bm{\Lambda}$ and $\bm{R\Lambda}$), 
we have three possibilities for the input size $Q$ (5, 10 or 15 NS) and three noise levels (see Table \ref{tab:Sets}), corresponding to a total of 27 different models. Where for these 27 different models we always have the same  $\mathbb{D}$ size, which is $\mathbb{D}=20 \times 32058 $ for training and  $\mathbb{D}=1 \times 3562 $ for testing.\\

The three levels of input noise are the following: i) a noiseless case, $\mathbb{X}_0$; ii) a small amount of noise measured by $\sigma_{M^R } =0.1 M_\odot$, $\sigma_R=0.2$ km, and $\sigma_{\Lambda}( M^\Lambda_q )=0.5\hat{\sigma}(M^\Lambda_q)$, $\mathbb{X}_1$; and iii) a larger noise level compatible with present observations, given by $\sigma_{M^R }=0.136 M_\odot$, $\sigma_R= 0.626$ km, where the process to obtain this values is described in our previous works \cite{PhysRevD.109.123038, Carvalho:2024zyb}, and $\sigma_{\Lambda}( M^\Lambda_q )=2\hat{\sigma}(M^\Lambda_q)$. We have chosen to characterize the tidal deformability noise by the standard deviation of the training dataset $\hat{\sigma}(M^\Lambda_q)$, e.g., $\hat{\sigma}(1.4M_{\odot})= 130.05$.  
The primary goal of creating these three distinct datasets is to evaluate the model's performance under varying noise levels.
 \\

\begin{table}[h!]
    \caption{The parameters used for generating the mock data for each set. The interpolation function $\hat{\sigma}(M_q^{\Lambda})$ was derived from the training dataset and represents the standard deviation for each possible value of mass.}
    \centering
      \setlength{\tabcolsep}{5pt}
\renewcommand{\arraystretch}{1.2}
    \begin{tabular}{ccccc}
  \hline
  \hline
    Set $\mathbb{X}$ &$\sigma_{ M^R } \;[ M_\odot]$ &$\sigma_R \;$ [km]  & $\sigma_{ M^\Lambda } \;[ M_\odot]$ & $\sigma_{\Lambda}( M^\Lambda_q )$ \\ \hline
    $R_0$  & 0     & 0& - & -  \\ \hline
    $R_1$  & 0.1  & 0.2&  - & - \\ \hline 
    $R_2$  &  0.136   & 0.626&  - & -  \\ \thickhline
    $\Lambda_0$  & - & - & 0     & 0  \\ \hline
    $\Lambda_1$  &  - & - & 0 & $0.5\hat{\sigma}(M^\Lambda_q)$  \\ \hline
    $\Lambda_2$  &   - & - &0  & $2\hat{\sigma}(M^\Lambda_q)$  \\ 
    \thickhline
        $R\Lambda_0$  &0     & 0& 0     & 0  \\ \hline
    $R\Lambda_1$  & 0.1  & 0.2& 0 & $0.5\hat{\sigma}(M^\Lambda_q)$  \\ \hline
    $R\Lambda_2$  &    0.136   & 0.626 &0  & $2\hat{\sigma}(M^\Lambda_q)$  \\  
    \hline
  \hline
    \end{tabular}
    \label{tab:Sets}
\end{table}

\section{Neural Networks \label{NN}} 

The objective of this study is to apply NNs to classify NS observations, distinguishing between those that suggest the presence of hyperons in the NS composition and those that do not. This classification will be performed across datasets with varying input noise levels and sizes, as explained in the previous section. Additionally, we will explore the impact of incorporating tidal deformability into the classification process. In this section, we will delve into the mechanics of NNs and their specific application to our research.

\subsection{How it is defined}
A NN consists of interconnected neurons organized into layers, including input $s=0$, hidden $s=1, ..., S-1$ and output layers, $s=S$, where $S+1$ is the total number of layers. The data is composed of a set of $(\bm{x}^{(i)},\bm{y}^{(i)})$ tuples, i.e.,  $D=\{ (\bm{x}^{(i)},\bm{y}^{(i)})\}^{\mathcal{D}}_{i=1}$, where $\bm{x}^{(i)} \in \mathbb{R}^Q$ and $\bm{y}^{(i)} \in \mathbb{R}^K$. Each layer is composed of neurons, which are connected to neurons in adjacent layers through weights, denoted as the matrix $\bm{W}$. 
Additionally, each neuron has a bias term, denoted as $\bm{b}$, which serves as a threshold. The network output is defined as $\hat{\bm{y}}= \text{NN}_{\bm{\theta}}(\bm{x})$, which is parameterized by $\bm{\theta}=\{(\bm{W}^{\pr{s}},\bm{b}^{\pr{s}})\}_{s=1}^S$. 
The computation within each neuron involves multiplying the weights with the corresponding neurons from the previous layer and summing these products for each hidden unit. The bias term is also added to the sum, giving a result of $\bm{\Sigma}= \bm{W} \bm{x}+ \bm{b}$.
Subsequently, a non-linear function, known as the activation function, and denoted by $\phi(\cdot)$, is applied to each neuron within both the hidden layers and the output layer. One of the most commonly used activation functions is the Rectified Linear Unit (ReLU) \cite{nair2010rectified}, defined as $\phi(x) = \max(0, x)$. This activation function determines the neuron's output, essentially dictating how "active" or responsive the neuron becomes. The final output for the $n$th neuron is therefore expressed as $\bm{a} = \phi(\bm{\Sigma})$. For the first hidden layer, the output of the hidden units is given by

\begin{equation}
\begin{bmatrix}
     \text{a}_{11}^{(i)} \\
     \vdots \\
     \text{a}_{n1}^{(i)}
 \end{bmatrix}
 = \phi^{\pr{1}} \pc{
\begin{bmatrix}
W_{11}^{\pr{1}} & \cdots & W_{1q}^{\pr{1}} \\
   \vdots & \ddots & \vdots \\
   W_{n1}^{\pr{1}} & \cdots & W_{nq}^{\pr{1}}
 \end{bmatrix}
\begin{bmatrix}
     x_1^{(i)} \\
     \vdots \\
     x_q^{(i)}
 \end{bmatrix}
+
\begin{bmatrix}
     b_{1}^{\pr{1}} \\
     \vdots \\
     b_n^{\pr{1}}
 \end{bmatrix}} \;.
\end{equation}

\noindent
\noindent Passing through all layers, the final result is calculated as

\begin{multline}
\text{NN}_{\bm{\theta}}(\bm{x})= \phi^{\pr{S}}\Biggl(\bm{W}^{\pr{S}} \underbrace{\phi^{\pr{S-1}}  \pc{\cdots\phi^{\pr{1}}\pc{\bm{W}^{\pr{1}}\bm{x} +\bm{b}^{\pr{1}}}\cdots }}_{\textbf{a}_{S-1}} \\ +\bm{b}^{\pr{S}}\Biggr)\;,
\end{multline}

\noindent in matrix form, this becomes
\begin{equation}
\begin{bmatrix}
     \hat{y}_{1}^{(i)} \\
     \vdots \\
     \hat{y}_{k}^{(i)}
 \end{bmatrix}
 = \phi^{\pr{S}} \pc{
\begin{bmatrix}
W_{11}^{\pr{S}} & \cdots & W_{1h}^{\pr{S}} \\
   \vdots & \ddots & \vdots \\
   W_{k1}^{\pr{S}} & \cdots & W_{kh}^{\pr{S}}
 \end{bmatrix}
\begin{bmatrix}
\text{a}_{1S-1}^{(i)} \\
     \vdots \\
     \text{a}_{hS-1}^{(i)}
 \end{bmatrix}
+
\begin{bmatrix}
     b_{1}^{\pr{S}} \\
     \vdots \\
     b_k^{\pr{S}}
 \end{bmatrix}} \;.
\end{equation}
Note that in the present work, while the input vector $\bm{x}^{(i)}$ has different dimensions, depending on the dataset type or number of NS being considered (see Table \ref{tab:Sets}), e.g., $\bm{x}^{(i)} \in \mathbb{R}^{2Q}$, for $\bm{R}$ and $\bm{\Lambda}$ and  $\bm{x}^{(i)} \in \mathbb{R}^{4Q}$ for $\bm{R\Lambda}$ with $Q=\{5,10,15 \}$, the output dimension is always fixed to one, as we are dealing with a single probability (scalar), i.e., $\bm{y}^{(i)} \in \mathbb{R}^1$, meaning K=1.
In a typical multi-class classification problem the final activation function $\phi^{\pr{S}}$ consists of a softmax function \cite{goodfellow2016deep}, however for a binary classification problem is the sigmoid activation function, $\phi(x) = 1/{(1 + \exp(-x)})$, ensuring that the model's probability predictions, $\text{NN}_{\bm{\theta}}(\bm{x}^{(i)}) \equiv p$, is constrained between 0 and 1.

\subsection{How it is trained}\label{chap:NN_train}

The primary objective during training is to minimize a designated loss function by optimizing the model's parameters, denoted as $\bm{\theta}$, in order to attain the lowest value of the loss
\begin{equation}
   \bm{\theta}^*= \argmin_{\bm{\theta}} L(\theta)\, .
\end{equation}
To achieve this, the back-propagation algorithm is employed, consisting of two main phases: the forward pass and the backward pass. In the forward pass, the model computes predictions and evaluates the loss function, which measures the error between the true output and the model's predictions.  For binary classification tasks, this is typically done using the binary cross-entropy loss function \cite{hinton2006reducing}, defined as
\begin{multline}
L(\bm{\theta})=\frac{1}{\mathcal{D}}\sum_{i=1}^\mathcal{D} \Biggl[\bm{y}^{(i)}\log(\text{NN}_{\bm{\theta}}(\bm{x}^{(i)}))  \\+(1-\bm{y}^{(i)})\log(1-\text{NN}_{\bm{\theta}}(\bm{x}^{(i)})) \Biggr]\,.
\end{multline}

\noindent The backward pass calculates derivatives of the loss function with respect to each parameter in the NN. These derivatives are then subtracted from the corresponding parameter values to update them
\begin{align}
\frac{\partial L(\bm{\theta})}{\partial \bm{\theta}}&=\frac{1}{\mathcal{D}}\sum_{i=1}^\mathcal{D} \frac{\partial l(\bm{y}^{(i)},\text{NN}_{\bm{\theta}}(\bm{x}^{(i)}))}{\partial \bm{\theta}} \;,\\
\bm{\theta'}&=\bm{\theta}-\eta\frac{\partial L(\bm{\theta})}{\partial \bm{\theta}} \, ,
\end{align} 
where $\eta$ represents the learning rate, a hyperparameter that governs the step size during parameter updates. The derivatives are computed using the chain rule
\begin{equation}
    \frac{\partial l(\bm{y}^{(i)},\text{NN}_{\bm{\theta}}(\bm{x}^{(i)}))}{\partial \bm{\theta}}= \frac{\partial l(\bm{y}^{(i)},\text{NN}_{\bm{\theta}}(\bm{x}^{(i)})) }{\partial \text{NN}_{\bm{\theta}}(\bm{x}^{(i)})} \frac{\partial \text{NN}_{\bm{\theta}}(\bm{x}^{(i)}) }{\partial \phi^{\pr{S}} } \frac{\partial \phi^{\pr{S}} }{\partial \bm{\theta}} \;.
\end{equation}
Weight updates in NNs are commonly accomplished through minibatch gradient descent. It involves segmenting the dataset into smaller mini-batches, with the batch size determining the number of data points used to update the weights in each iteration. This method offers a strike balance between accurate optimizations and computation efficiency.
An epoch is then when all the training data has been used.

\subsection{Training procedure for our problem}

\begin{figure}[!hbt]
    \centering
\includegraphics[width=1\linewidth]{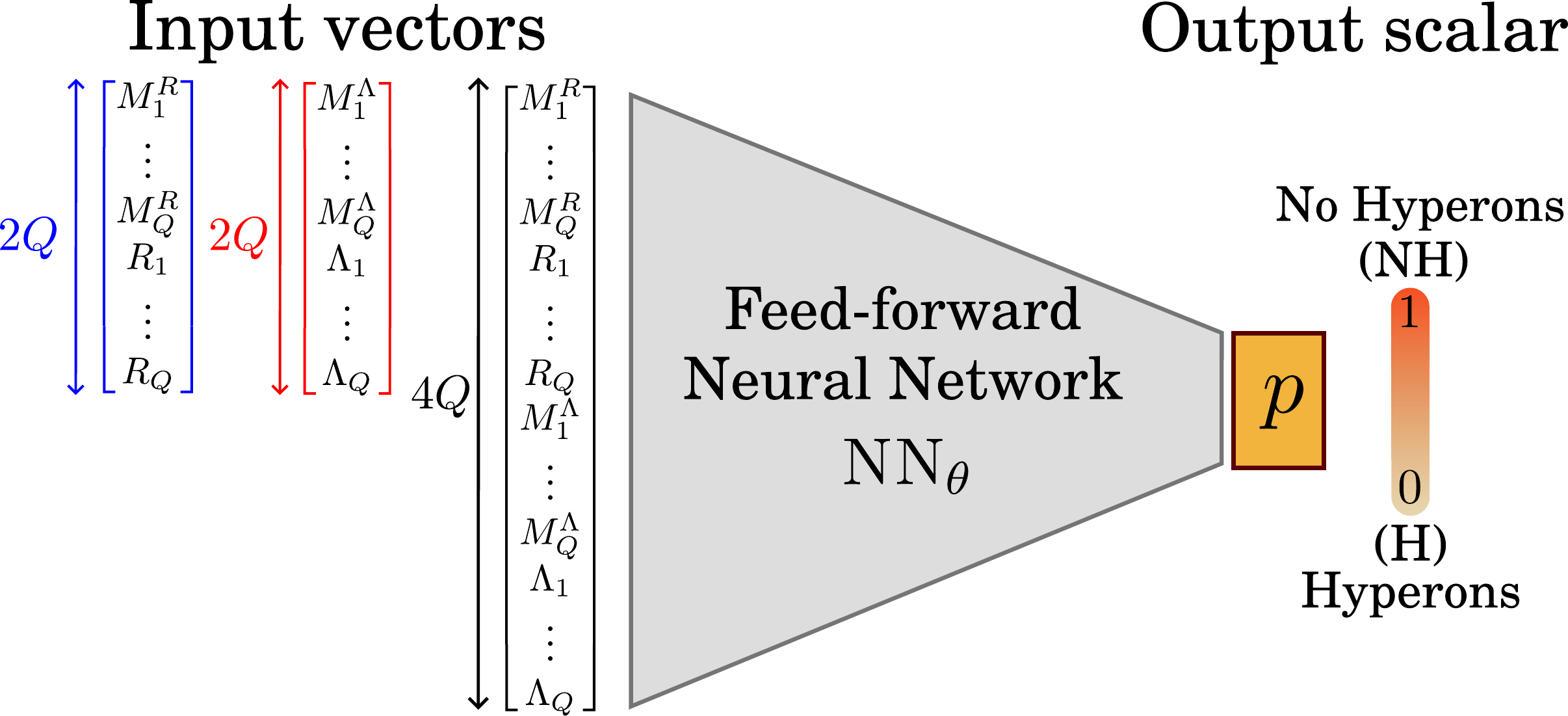}
    \caption{Schematic representation of the classification model. Depending on the type of model, the size of the input vectors is $2Q$ ($R_i$ and $\Lambda_i$ models) and $4Q$ ($R\Lambda_i$ model), where $Q$ indicates the number of NS used (5, 10 and 15). The output of the model is a scalar value $p \in [0,1]$, where the value 0 identifies the hyperons class whereas 1 indicates the absence of hyperons (no hyperons class). }
    \label{fig:drawing}
\end{figure}

A schematic representation of the problem is present in Fig.~\ref{fig:drawing}, where the three different input types, each associated with three distinct noise levels, are represented (see Table \ref{tab:Sets}). Additionally, each input type varies by the number of stars used (5, 10, or 15). The output is a scalar value, defined as a probability distribution, $p$, where we assign the class {\it No Hyperons} (NH) as 1 and the class {\it Hyperons} (H) as 0.
For the training procedure, a random subset of the training data was reserved for validation, with an 80/20 split for training and validation, respectively. The input data $\mathbb{X}$ was standardized.\\

The optimal machine learning model is identified through the tuning of its hyperparameters, including the number of neurons, layers, and activation functions. For the hidden layers, we explored ReLU, Softplus, and Sigmoid activation functions, with a Sigmoid activation function for the output layer. For the sets $\bm{R}$ and $\bm{\Lambda}$, which share the same input size, various NN architectures were tested. A grid search revealed that all input configurations performed equally well with the same architecture, see Table \ref{tab:final_model_R}.

For the $\bm{R\Lambda}$ sets, a separate grid search was performed to identify the most effective model architectures for each of the three input sizes. The results, presented in Table \ref{tab:final_model_RL}, show that the optimal architecture varies with input size. This suggests that as the number of input features increases 

and the data becomes more complex—incorporating tidal deformability and mass-radius pairs—the model architecture must adapt accordingly.
For instance, with an input size of 20 ($Q=5$ stars), the model requires a greater number of layers, likely because the complexity of the input data demands a more sophisticated approach to capture the underlying patterns and interactions effectively than when more information is given for an input size of 40 ($Q=10$ stars).
One might wonder why the same architecture is not used for input sizes of 20 in the $\bm{R}$, $\bm{\Lambda}$, and $\bm{R\Lambda}$ sets. The answer lies in the higher complexity of the $\bm{R\Lambda}$ dataset, which combines information from both radius and tidal deformability, creating a more intricate problem for the model to solve. Consequently, the $\bm{R\Lambda}$ set requires a distinct architecture to handle this increased complexity, even when the input size is the same.
All grid searches were performed using set $\mathbb{X}_0$, with the assumption that similar behavior will be observed across the other sets.\\

During training, a linear scheduler was employed to gradually reduce the learning rate from 0.01 to 0.001. The ADAM optimizer \cite{kingma2014adam}, enhanced with the AMSgrad improvement \cite{reddi2019convergence}, was used. The models were trained for 2000 epochs, with the model achieving the lowest validation loss being selected, as early stopping was not implemented. The training process utilized a mini-batch size of 1024. The implementation was carried out using the \texttt{PyTorch} library \cite{NEURIPS2019_9015}.\\

\begin{minipage}{.5\textwidth}
      \centering
               \captionsetup{type=table}
  \caption{Parameters for Datasets $\bm{R}$ and $\bm{\Lambda}$.\label{tab:final_model_R}}
  \setlength{\tabcolsep}{5pt}
\renewcommand{\arraystretch}{1.2}
\begin{tabular}{ccc}
\toprule
\hline
\hline
\textbf{Layers} & \textbf{Activation}& \textbf{Neurons}\\ \hline
Input & N/A &{10,20,30} \\ \hline
Hidden Layer 1 & ReLU & 40\\ \hline
Hidden Layer 2 &  ReLU & 80\\ \hline
Hidden Layer 3 &  ReLU  & 40\\ \hline
Output & Sigmoid & 1 \\ \hline
\end{tabular}
\end{minipage}
\begin{minipage}{.5\textwidth}
         \centering
                  \captionsetup{type=table}
  \caption{Parameters for Datasets $\bm{R\Lambda}$.\label{tab:final_model_RL}}
    \setlength{\tabcolsep}{5pt}
\renewcommand{\arraystretch}{1.2}
\begin{tabular}{ccccc}
\toprule
\hline
\hline 
\textbf{Layers} & \textbf{Activation}& \multicolumn{3}{c}{\textbf{Neurons}}\\ \hline
Input & N/A &20&40&60 \\ \hline
Hidden Layer 1 & ReLU & 80&30&120\\ \hline
Hidden Layer 2 &  ReLU  & 100&2&120\\ \hline
Hidden Layer 3 &  ReLU & 80&-& -\\ \hline
Output & Sigmoid & \multicolumn{3}{c}{1} \\ \hline
\end{tabular}
\end{minipage}

\section{Results \label{results}}

After training the NN classification models, we evaluate their performance using several metrics. In this section, we describe these evaluation metrics and present the models' performance on an independent test set that shares the same theoretical framework as the training data. We then test the models on a dataset generated within a different theoretical framework to assess their robustness. Finally, we apply the trained models to real observational data, demonstrating their practical applicability.

\subsection{Metrics}

\begin{figure*}[!hbt]
    \centering
    \includegraphics[width=0.35\linewidth]{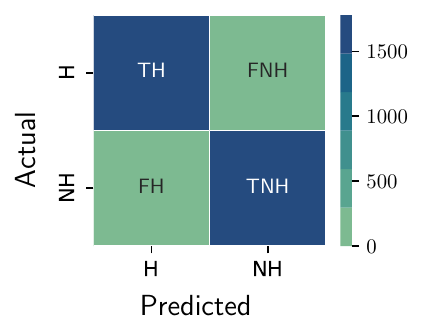}
    \includegraphics[width=0.5\linewidth]{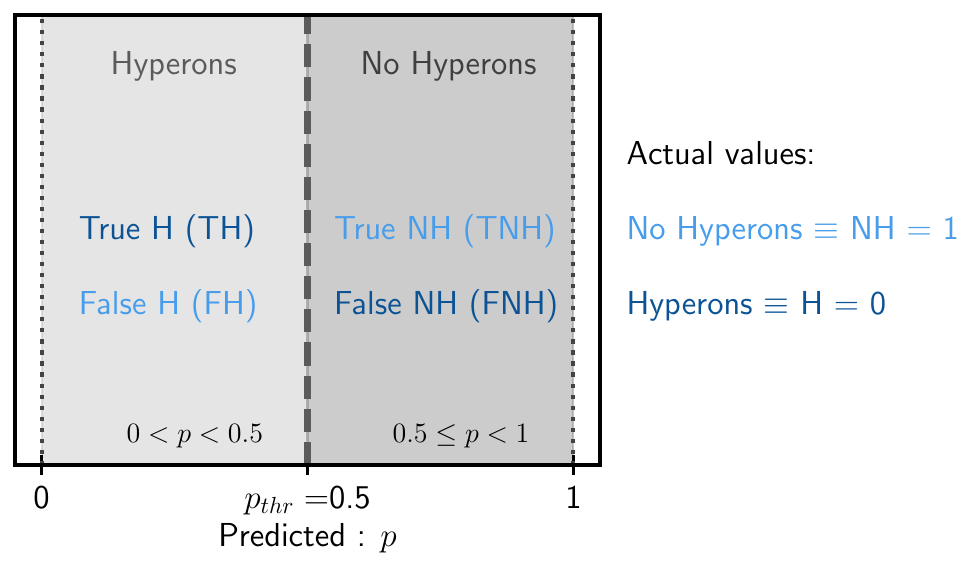}
    \caption{Left: The non-normalized confusion matrices for our specific classification task (detecting the presence of hyperons).  The size of the test sets is 1737 H samples and 1825 NH samples. The labels designate the four possible pairs of model input/prediction: True Hyperons (TH), True No Hyperons (TNH), False Hyperons (FH), and False No Hyperons (FNH). Right: Relation between the probability $p$ and the labels TH, TNH, FH and FNN. The threshold has been taken at $p=0.5$, which means that {\it Hyperons} ({\it No Hyperons}) corresponds to  $p<0.5 $  ($p\geq0.5)$.
   }
    \label{fig:ske}
\end{figure*}

To assess the performance of the probabilistic classifiers models, we convert the output probability estimates into binary classes. 
We defined that a given sample belongs to the class H if $p <0.5$ while the sample belongs to the class NH (only nucleons are present) if  $p \geq 0.5$. The models evaluations rely on key metrics derived from the confusion matrix, represented in Fig.~\ref{fig:ske}.
The confusion matrix offers a detailed overview of the model's injected (actual) samples vs. predicted outcomes, including the counts of True Hyperons (TH), True No Hyperons (TNH), False Hyperons (FH), and False No Hyperons (FNH). For instance, the value of TH gives the number of H samples correctly classified, while FNH denotes the number of wrongly classified H samples. These metrics provide a robust framework for analyzing the effectiveness and accuracy of our models.

The primary metric employed is accuracy, which represents the proportion of correctly classified instances (both TH and TNH) out of all instances. It is defined as
\begin{equation} \label{eq:ac} \text{Accuracy} = \frac{TH + TNH}{TH + TNH + FH + FNH} \;. \end{equation}
While accuracy is useful, it does not always provide a complete picture, especially when dealing with imbalanced datasets. To address this, we also use the F1 score, which is the harmonic mean of precision and recall, also known as True Positive Rate (TPR), where the precision is defined as 
\begin{equation} \label{eq:prec}
    \text{Precision} = \frac{TNH}{TNH+FNH}\; ,
\end{equation}
and the recall as 
\begin{equation} \label{eq:recall}
    \text{Recall} =\frac{TNH}{TNH + FH}\; .
\end{equation}
The F1 score offers a more balanced measure by considering both FH and FNH, and it is particularly effective for imbalanced classes. It is given by
\begin{equation} \label{eq:f1} \text{F1 score} = \frac{2 \times TNH}{2 \times TNH + FH + FNH} \;. \end{equation}

\subsection{Application to mock data}
\begin{figure*}[!hbt]
    \centering
    \includegraphics[width=0.33\linewidth]{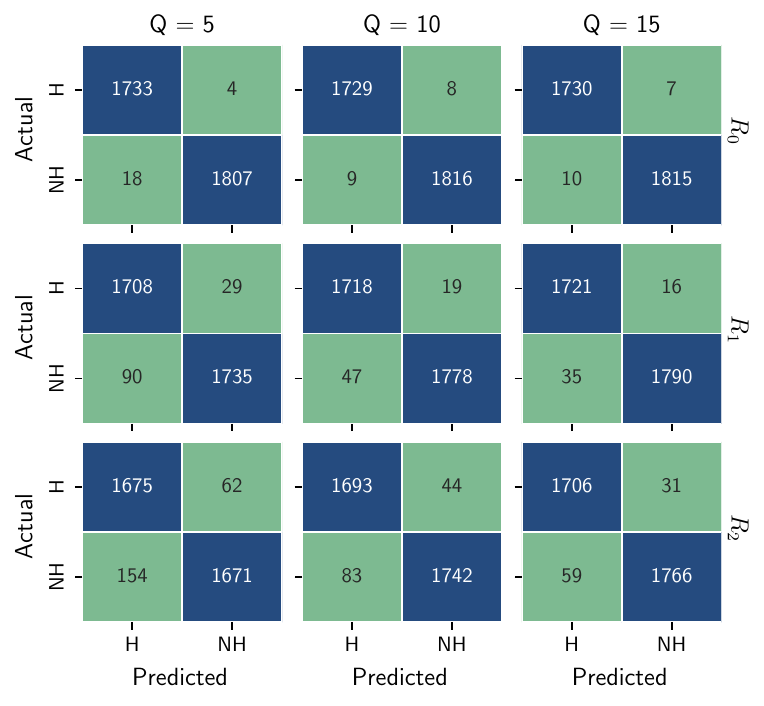}
    \includegraphics[width=0.33\linewidth]{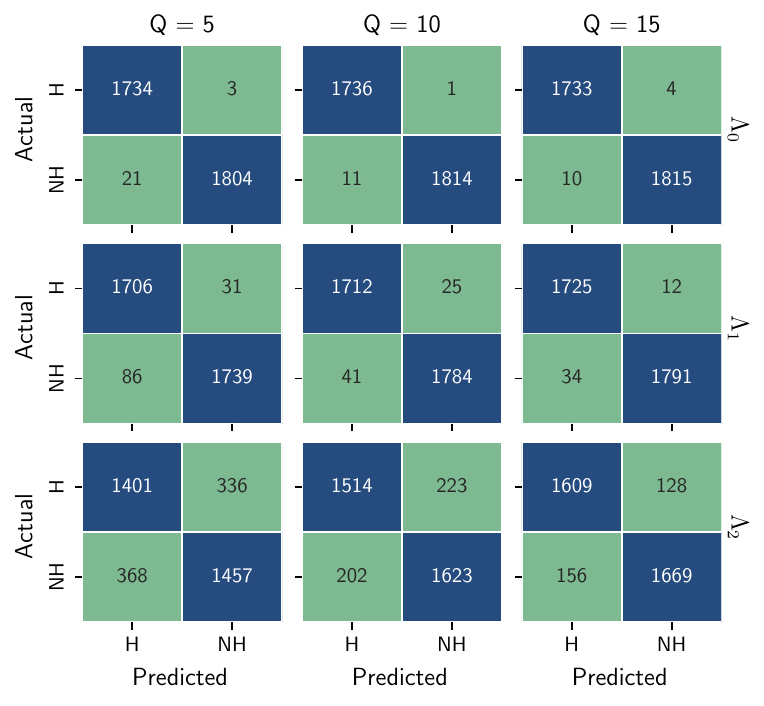}
    \includegraphics[width=0.33\linewidth]{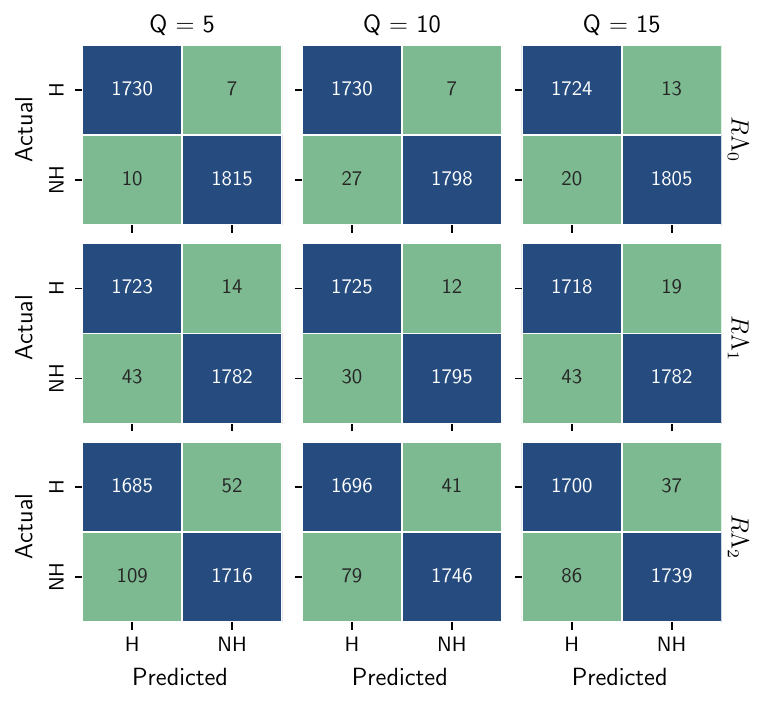}
    \caption{Non-normalized Confusion matrices for datasets using $\bm{R}$ (left), $\bm{\Lambda}$ (middle) and $\bm{R\Lambda}$  (right) inputs. The size of the test sets is 1737 H samples and 1825 NH samples. Each panel contains three columns for the different number of stars in each observation contains ($Q=5,\, 10,\, 15$)  and three rows for the different levels of noise considered $\mathbb{X}_i$, $i=0,1,2$ defined in Table \ref{tab:Sets}. }
    \label{fig:ML_c}
\end{figure*}

The confusion matrices for all models trained are show in Fig.~\ref{fig:ML_c}, which were computed on the test set. The different models, with architectures defined in Tab. \ref{tab:final_model_R} and \ref{tab:final_model_RL}, are denoted by the nature of their input data, meaning the data to which they were trained and the data to which they will be tested: $R_{i}$ for $R(M)$ data, $\Lambda_{i}$ for $\Lambda(M)$, and $R\Lambda_{i}$ when both $R(M)$ and $\Lambda(M)$ data are considered. The index $i=0,1,2$ denotes the three possible noise levels (see Table \ref{dataset}). A clear trend emerges from the analyses of the confusion matrices: as noise increases $i=0$ to $2$ across all input sizes and types, the counts of TH and TNH decrease. This trend indicates that the model's precision diminishes as noise levels rise, which is an expected outcome. 
The higher number of Actual NH instances observed in the confusion matrices is due to the random class distribution of the test set, which consists of 1737 H samples and 1825 NH samples. This test set distribution will be consistently used for evaluation purposes throughout the analysis. \\

For the $R_0$, $\Lambda_0$, and $R\Lambda_0$ (noiseless) models, the number of true classifications converges and ceases to improve once the input size reaches $Q=10$ NS. In the case of $\bm{R\Lambda}$, this convergence is observed across all noise levels. However, for $\bm{R}$ and $\bm{\Lambda}$ models, there is a noticeable improvement when increasing the number of NSs to 15 under noisy conditions. This suggests that when less information is provided to the model (as in the separate $\bm{R}$ and $\bm{\Lambda}$ inputs), more input NS are necessary to achieve a higher count of true classifications, especially in the presence of noise. Conversely, without noise, the model's precision reaches a plateau at $Q=10$ NS.\\\\
A detailed analysis of false predictions shows a substantial increase when comparing $\mathbb{X}_0$ and $\mathbb{X}_2$, which can be quantified by the ratio $\mathbb{X}_2/\mathbb{X}_0$, where we are using $(FH(\mathbb{X}_2) + FNH (\mathbb{X}_2))/(FH(\mathbb{X}_0) + FNH (\mathbb{X}_0))$. For the $\bm{R\Lambda}$ models, this ratio is 9.5, 3.5, and 3.7 from $R\Lambda_0$ to $R\Lambda_2$ for $Q = 5$, $Q = 10$, and $Q = 15$, respectively. In comparison, the $R$ models exhibit ratios of 9.8, 7.5, and 5.3 for the same values of $Q$. The $\bm{\Lambda}$ models, however, display significantly higher ratios, with false predictions increasing by 29.3, 35.4, and 20.3 for $Q = 5$, $Q = 10$, and $Q = 15$, respectively.
These differences highlight varying sensitivities to the number of samples across different sets. The $\bm{\Lambda}$ models exhibit a significantly higher increase in false values compared to the $\bm{R}$ and $\bm{R\Lambda}$ models, leading to a more pronounced decline in performance. In contrast, the $\bm{R\Lambda}$ models demonstrate greater stability, with less fluctuation in false values and maintaining a more consistent performance across different sample sizes.

\begin{figure}[!hbt]
    \centering
\includegraphics[width=0.9\linewidth]{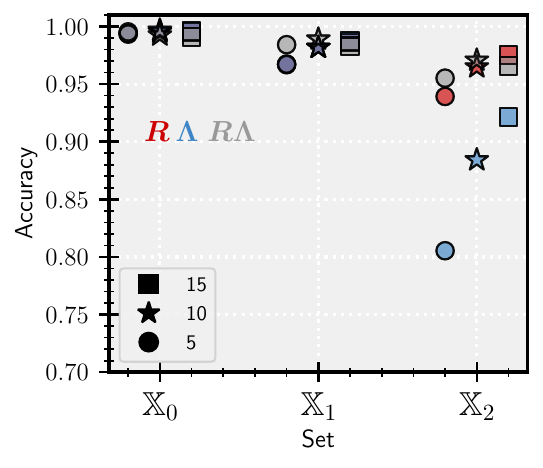}
    \caption{Accuracy for all dataset types, $\bm{R}$ (in red), $\bm{\Lambda}$ (in blue) and $\bm{R\Lambda}$  (in gray) (see Table \ref{dataset}). The x-axis indicates the noise size $\mathbb{X}_i,\, i=0,1,2$ (see Table \ref{tab:Sets}) and the symbols denote the number of NSs of each observation ($Q=5, 10, 15$).}
    \label{fig:accuracy}
\end{figure}

{Let us analyze the behaviour of the Accuracy, Eq.~(\ref{eq:ac}),  over the  different datasets.   As can be observed in Fig. \ref{fig:accuracy}, accuracy increases progressively as the noise level decreases from 2 to 0 (noiseless) across all type models ($R_i$, $\Lambda_i$, and $R\Lambda_i$). The impact of varying the number of NS in each observation (denoted by Q) is best analyzed in Tab. \ref{tab:accuracy_values}, where we also present the F1 score and the Area Under the Curve (AUC) for accuracy as the threshold varies.}

The F1 score is a consequence of the classes representing the presence (H class) or absence (NH class) of hyperons being equally weighted - implying that the models treat false H and false NH with equal importance. As is typical in binary classification tasks, we have set a probability threshold $p_{\text{thr}}$ of $0.5$ for the binary classification: a sample belongs to the H class if $p <0.5$ otherwise it belongs to the NH class ($p \geq 0.5$).
To discuss the consequences of choosing a different probability threshold $p_{\text{thr}}$ on the performance of the models, we introduce the concept of the AUC of the model's accuracy as the threshold $p_{\text{thr}}$ varies.
While AUC is commonly associated with the Receiver Operating Characteristic (ROC) curve, as discussed in \cite{kim2015hybrid}, we use it here to evaluate the accuracy of the model across different thresholds. This approach allows us to measure how well the model performs independently of a particular threshold. An AUC of 1 would indicate perfect classification. 

The highest accuracy values, see Tab. \ref{tab:accuracy_values}, although differing by only 0.001, are observed with $Q=10$ for both $R_0$ and $\Lambda_0$. For $R\Lambda_0$ a higher accuracy is obtained with a smaller number of pairs ($Q=5$), probably due to the additional information provided by the combination of mass-radius and mass-tidal deformability inputs.
For the $R_1$ to $R_2$ and $\Lambda_1$ to $\Lambda_2$ models, accuracy increases steadily as the number $Q$ of pairs increases, with more pronounced improvements observed in higher noise scenarios, where each additional pair significantly improves the performance of the model. In contrast, the accuracy for $R\Lambda_1$ and $R\Lambda_2$ tends to plateau at $Q=10$, indicating that beyond this point, adding more pairs does not improve accuracy further, possibly due to the model reaching its capacity to effectively process the combined data.
The F1 score results show a very similar behaviour to the accuracy results, and we can see that they only differ by the third decimal place for some values.
The AUC value shows the same behaviour as the Accucay in the majority of cases.  This consistency suggests that the results are not very sensitive to the choice of threshold.

\begin{table*}[!hbt]
\caption{Accuracy, F1 score and AUC value across the multiple models, as plotted in Fig. \ref{fig:accuracy}.}

\centering
\setlength{\tabcolsep}{5pt}
\renewcommand{\arraystretch}{1.1}
\begin{tabular}{c|cccccccccccc}
\hline 
\hline 
\multirow{2}{*}{\diagbox[]{Set}{Q}}& &\multicolumn{3}{l}{Accuracy}& &\multicolumn{3}{l}{F1 score} &\multicolumn{3}{c}{AUC}  \\
& 5 & 10 & 15&$\phantom{a}$ &5 & 10 & 15 &$\phantom{a}$&5 & 10 & 15 \\ \hline 
$R_0$& 0.994 & 0.995 & 0.995& & 0.994 & 0.995 & 0.995 && 0.977 & 0.979 & 0.980 \\
$R_1$ & 0.967 & 0.981 & 0.986& & 0 0.967 & 0.982 & 0.986& & 0.927 & 0.953 & 0.965\\
$R_2$& 0.939 & 0.964 & 0.975& & 0.939 & 0.965 & 0.975& & 0.872 & 0.916 & 0.940 \\ \thickhline 
$\Lambda_0$& 0.993 & 0.997 & 0.996& & 0.993 & 0.997 & 0.996& & 0.978 & 0.983 & 0.981 \\
$\Lambda_1$& 0.967 & 0.981 & 0.987& & 0.967 & 0.982 & 0.987& & 0.920 & 0.951 & 0.965 \\
$\Lambda_2$ &0.802 & 0.881 & 0.920& & 0.805 & 0.884 & 0.922& & 0.642 & 0.765 & 0.842 \\ \thickhline 
$R\Lambda_0$ & 0.995 & 0.990 & 0.991 & & 0.995 & 0.991 & 0.991  && 0.981 & 0.970 & 0.973\\
$R\Lambda_1$  & 0.984 & 0.988 & 0.983& & 0.984 & 0.988 & 0.983  && 0.956 & 0.956 & 0.960 \\
$R\Lambda_2$  & 0.955 & 0.966 & 0.965& &  0.955 & 0.967 & 0.966 && 0.905 & 0.914 & 0.927 \\\hline \hline
\end{tabular}
\label{tab:accuracy_values}
\end{table*}

Overall, our models demonstrate strong performance across all evaluated sets, with consistently high metric values, particularly in terms of accuracy and F1 score. Although there is a noticeable decline in precision as noise is introduced - especially when examining the predictions for $\mathbb{X}_2$, which contains noise levels similar to those found in actual observations - the models still maintain a high level of accuracy. This is particularly evident in the $\bm{R}$ and $\bm{R\Lambda}$ models, where accuracy remains robust despite the added noise.
These results are encouraging as they suggest that our models are well-equipped to handle real-world observational data with inherent noise.

\subsection{Application to different hadronic models \label{DDB}}

\begin{figure}[!hbt]
    \centering
    \includegraphics[width=0.9\linewidth]{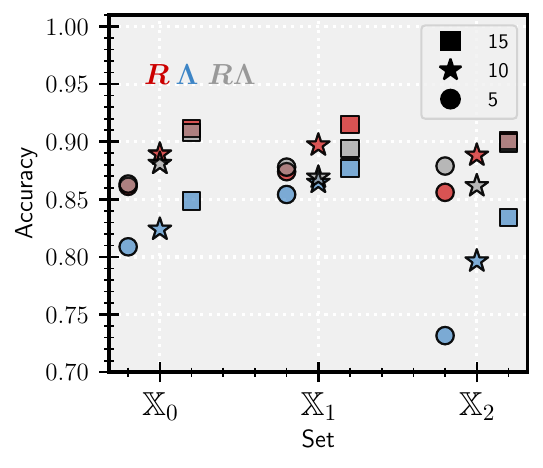}
     \caption{Accuracy for the new dataset, analogous to Fig. \ref{fig:accuracy}, showing results for $\bm{R}$ (red), $\bm{\Lambda}$ (blue), and $\bm{R\Lambda}$ (grey) (see Table \ref{dataset}). The x-axis represents the noise size $\mathbb{X}_i$ where $i = 0, 1, 2$ (refer to Table \ref{tab:Sets}). Symbols on the plot indicate the number of NSs for each observation, corresponding to $Q = 5$, $10$, and $15$.
    }

    \label{fig:accuracy2}
\end{figure}

Having confirmed the ability of our model to predict the test set, we extend our work to evaluate its performance with additional data sets generated within a different description of nuclear matter. Specifically, we consider the sets  DDB from \cite{providencia2023neutron} with 19140 EoS with hyperons and 8794 EoS without hyperons. These sets have been calculated within an RMF framework with density dependent couplings \cite{Typel1999,Typel2009,Malik:2022zol}.
To test the model on this new EoS data, we have created corresponding datasets for each model, incorporating the same levels of noise as defined in Tab. \ref{tab:Sets}, following the procedure outlined in section \ref{dataset} for the test set with $n_s=1$. This resulted in a combined dataset of $\mathbb{D}=27934$ samples.
The results for accuracy are presented in Fig. \ref{fig:accuracy2}, and Tab \ref{tab:accuracy_values2} provides a detailed breakdown of the accuracy, F1 score, and AUC values. The overall accuracy remains consistently above 73\%, demonstrating that the model reliably predicts the results. Notably, accuracy increases with the number of observations ($Q$) for most sets, except for a slight dip from $Q=5$ to $Q=10$ in sets $R\Lambda_1$ and $R\Lambda_2$.
The higher accuracy observed for the set $\mathbb{X}_1$ compared to $\mathbb{X}_0$ can be explained by the regularizing effect of noise, which helps prevent overfitting and improves the model's ability to generalize to unseen data, i.e. adding a small amount of noise acts as a regularizer, preventing the model from fitting too closely to the specific details of the training data. In the absence of noise, the model may overfit by focusing too much on small patterns, reducing its generalizability.
Specifically, the DDB sets predict a larger radius and tidal deformability upper limit for nuclear matter, as well as a smaller lower limit for both quantities in the presence of hyperons, compared to the model’s training set, defined as NL. The observed decline in accuracy for $R\Lambda$ at $Q=10$ relative to $Q=5$ likely comes from increased discrepancies in the joint probability of $R\Lambda$, reflecting formalism differences alongside the uncertainties in set $\mathbb{X}_1$. 
For set $\mathbb{X}_2$, which includes larger uncertainties,  the expected trend resumes, with accuracy increasing as the number of observations, $Q$, grows. Also, we should be aware that maybe some slight overfitting can be happening here given that for Q=10 which now has the smaller performance in the previous section had the biggest. \\

Looking at Tab. \ref{tab:accuracy_values2}, given the relatively small proportion of NH, it is important to consider not only accuracy but also the F1 score. The F1 score closely mirrors the accuracy trend, reaching values of 0.876, 0.873, and 0.871 for the top-performing sets ($R_0$, $R_1$, $R\Lambda_0$) at $Q=15$, suggesting the model effectively handles class imbalances and difficult-to-classify samples. In addition, the AUC values, which quantify the model’s ability to distinguish between classes independently of the classification threshold, range from 0.534 (for $\Lambda_2$ at $Q=5$) to 0.867 (for $R_0$ at $Q=15$). Most $R$ and $R\Lambda$ sets achieve AUC values above 0.8 at higher $Q$ levels, indicating strong discriminatory power, particularly when more observations (NSs) are available.

\begin{table*}[!hbt]
\caption{Accuracy, F1 score and AUC value across the multiple models, as plotted in Fig. \ref{fig:accuracy2} for the new set.}
\centering
\setlength{\tabcolsep}{5pt}
\renewcommand{\arraystretch}{1.1}
\begin{tabular}{c|cccccccccccc}
\hline 
\hline 
\multirow{2}{*}{\diagbox[]{Set}{Q}}& &\multicolumn{3}{l}{Accuracy}& &\multicolumn{3}{l}{F1 score} &\multicolumn{3}{c}{AUC}  \\
& 5 & 10 & 15&$\phantom{a}$ &5 & 10 & 15 &$\phantom{a}$&5 & 10 & 15 \\ \hline 
$R_0$&  0.861 & 0.889 & 0.911 & & 0.817 & 0.849 & 0.876 & &0.807 & 0.838 & 0.867 \\
$R_1$ & 0.876 & 0.897 & 0.911 &  & 0.820 & 0.855 & 0.873 & & 0.785 & 0.823 & 0.852 \\
$R_2$& 0.853 & 0.884 & 0.897 & & 0.784 & 0.834 & 0.854 & &0.735 & 0.790 & 0.818 \\ \thickhline 
$\Lambda_0$& 0.809 & 0.824 & 0.849 && 0.765 & 0.781 & 0.806 & &0.751 & 0.774 & 0.802 \\
$\Lambda_1$& 0.854 & 0.865 & 0.877 & & 0.798 & 0.818 & 0.834 & &0.751 & 0.781 & 0.808 \\
$\Lambda_2$ &0.732 & 0.796 & 0.834 & & 0.652 & 0.733 & 0.779 & &0.534 & 0.645 & 0.707  \\ \thickhline 
$R\Lambda_0$ & 0.863 & 0.881 & 0.908 & & 0.821 & 0.839 & 0.871 & &0.812 & 0.809 & 0.863 \\
$R\Lambda_1$  & 0.878 & 0.869 & 0.894 & & 0.832 & 0.805 & 0.853 & &0.803 & 0.804 & 0.833\\
$R\Lambda_2$  & 0.879 & 0.862 & 0.899  & & 0.826 & 0.772 & 0.853  & &0.780 & 0.792 & 0.810  \\\hline \hline
\end{tabular}
\label{tab:accuracy_values2}
\end{table*}

Overall, the models $\bm{R}$ and $\bm{R\Lambda}$ demonstrated superior performance, while the model struggled the most with $\Lambda_2$, which exhibited lower accuracy, F1-scores, and AUC values. This suggests that the dataset used to train the model contains excessive noise, making it harder for the model to capture patterns effectively. In the case of $\Lambda_2$, the absence of radius information, unlike in $R\Lambda_2$, further complicates the model's ability to learn the underlying data behavior.
In summary, while the model performs well across most datasets, there is room for improvement, particularly in handling the $\Lambda$ datasets with higher noise levels. In the following sections, we will investigate the model’s performance using real observational data.

\subsection{Application to real data}

\begin{figure}[!hbt]
    \centering
\includegraphics[width=1.01\linewidth]{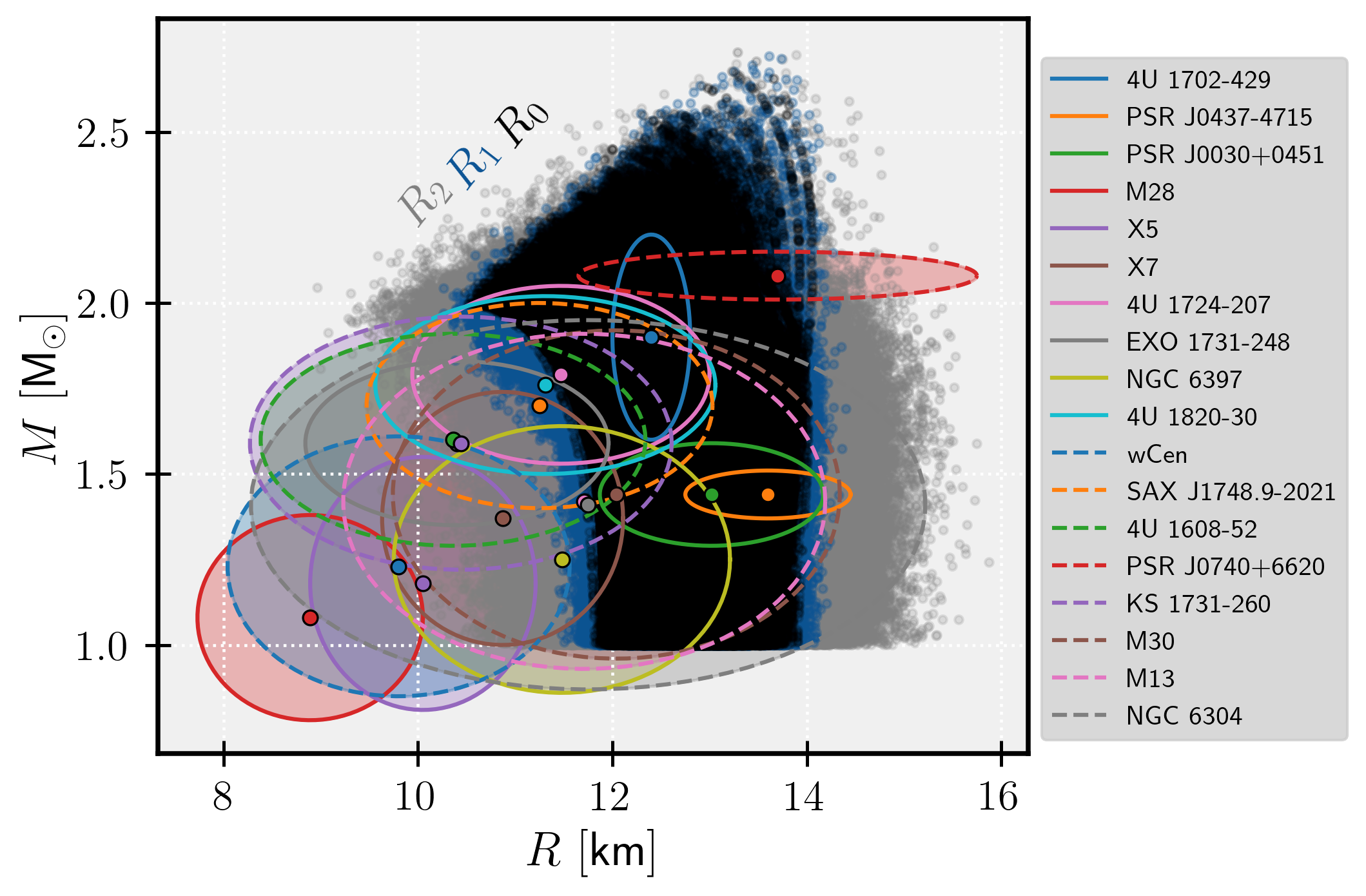}
    \caption{Observations taken from Table 1 of \cite{Soma:2022vbb}, for the 68\% CI, the gray, blue, and black bands represent the $R_2$, $R_1$, and $R_0$ data sets, respectively, and include both solutions with and without hyperons.
    }
    \label{fig:Obs}
\end{figure}

As a final test of our classification models, we apply the models that use $R(M)$ as input data (denoted as $\bm{R}$ in Table \ref{dataset}) to real NS data. Although the use of tidal deformability data obtained from GW observations would be ideal for constraining the EOS and searching for evidence of hyperons, the current lack of sufficient observational data limits this approach. However, future advances in GW detectors, such as the next generation of ground- and space-based observatories, are expected to play a critical role in this area. 
Given the model's limited extrapolation capabilities, we carefully selected a subset of NSs observations, listed in Table 1 of \cite{Soma:2022vbb}, ensuring they fell within the model's training region. 
For the $R_0$ and $R_1$ models, we have selected nine observations where the central values of the radius distributions were within $\pm 0.2$ km of the training data range, as shown in Fig.   \ref{fig:Obs}.

\begin{figure}[!hbt]
    \centering
\includegraphics[width=0.9\linewidth]{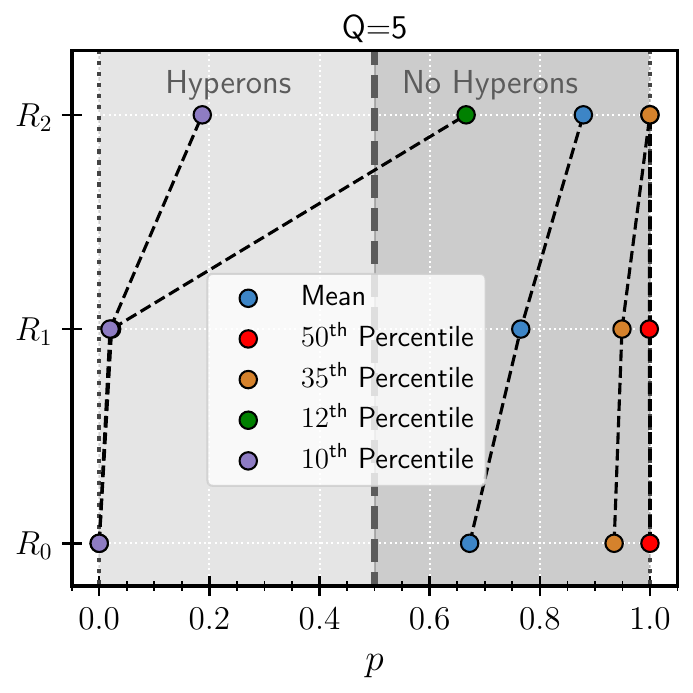}
    \caption{Some statistics (mean, median, 35$^{\text{th}}$, 12$^{\text{th}}$ and 10$^{\text{th}}$ percentiles) for the $R_2$, $R_1$, and $R_0$ models represented in the y axis, for $Q=5$. The probability $p=0.5$ was considered as the threshold to define the labels {\it Hyperons} and {\it No Hyperons}. }
    \label{fig:prevision}
\end{figure}

For $R_2$ (the trained model with a larger noise value), which better matches the dispersion of the observational data (as illustrated in Fig.  \ref{fig:Obs}), we included all observations except \textit{M28}. This allowed model $R_2$ to exhibit an input noise level similar to the standard deviation of the mass-radius pairs, providing a more realistic test scenario. To minimize extrapolation given the broad distribution within 1$\sigma$ for each observation, we opted to use the mean values of the observations as inputs to the NN models.
To generate test data, we employ sampling without replacement to generate 100 samples of the input vector of five NS out of the 9 observations for models $R_0$ and $R_1$ and 17 for $R_2$. The statistics of the models outputs are shown in 
 Fig. \ref{fig:prevision}, which displays the mean and the more relevant percentiles for the output predictions. This figure is based on Fig. \ref{fig:ske} and also considers the probability $p=0.5$ as the threshold to define the labels {\it Hyperons} and {\it No Hyperons}.
The results indicate that although the model produces predictions across the whole spectrum, the 35$^{\text{th}}$ percentile falls in the {\it No Hyperons} range completely across all three models. This implies that 65\% of the data across the three sets lies in the {\it No Hyperons} region. 
In addition, the median is completely at $p=1$, suggesting that the observational data are more consistent with NSs lacking hyperons. Note that the confidence of the {\it No Hyperons} case is larger for the $R_2$ model, as can be clearly seen especially for the 12$^{\text{th}}$ percentile, where it is the only model that already falls within the range indicating the absence of hyperons. These results are especially meaningful because $R_2$ is the model trained with data that closely matches the spread observed in NS observations, thereby reinforcing the reliability of this outcome.

We further tested the $R_2$ model using the model with $Q = 10$ and $Q = 15$, considering the 17 available observations. Due to the limited number of values available for varying the input vectors, we conducted 10 samplings without replacement, resulting in 10 different input vectors. For $Q = 15$, all predictions were equal to 1, and for $Q = 10$, all predicted class probabilities were $p > 0.896$. These results further validate our earlier conclusion, especially given the high scores achieved for the metrics with $Q=10$ and 15 in the previous section, indicating the reliability of our model's predictions.

{
To the best of our knowledge, our work represents the first attempt to frame the detection of hyperons as a classification problem. However, there have been attempts to identify the characteristics which define the properties of NSs that contain hyperons in their composition. For instance, in the study \cite{fortin2015neutron}, where the authors study 14 theoretical EOS of dense matter based on the RMF framework, and analyse the effects of the presence of hyperons, 
they conclude that medium mass stars with hyperons have larger radii than pure nucleonic stars with the same mass. This is also a feature of the datasets we have used to define our classification models.   
In another study,  \cite{sun2023astrophysical}, the authors study the implications of considering the presence of hyperons within NS on the hyperon couplings. They apply a Bayesian inference calculation to constrain the hyperon couplings from hypernuclear properties and astrophysical observations of NS, and conclude that the former are much more constraining than the astrophysical data. This seems to indicate that in order to identify hyperons inside NS measurements with smaller uncertainties are needed.
Another potential indicator of hyperon presence in NSs comes from the cooling curves. In a model that does not predict nucleonic direct Urca, because the proton fraction remains too small in the NS interior, the opening of direct Urca processes may occur only after the hyperon onset, allowing rapid cooling in the NS \cite{Providencia:2018ywl,Fortin:2020qin}. However, the cooling curves are subject to much larger uncertainties than the NS radii, and therefore the information we can extract from the cooling curves is still limited.
}

\section{Conclusions \label{conclusion}}

We investigated a classification problem aimed at determining the potential presence of hyperons within NSs based on their macroscopic properties. We investigate the capacity of NNs for the present binary classification problem of detecting the presence of hyperons.  
The dataset used has been constructed from an RMF approach within a Bayesian framework, incorporating crucial constraints derived from both nuclear matter properties and NS observations. The dataset consists of 17810 EOSs for pure nucleonic matter and 18728 EOSs including hyperons.
This approach was chosen to achieve our primary objective: to investigate the presence of hyperons within NSs based on observational data, which requires the use of microscopic models. 
From this comprehensive dataset, we generated three distinct sets by selecting either 5, 10, or 15 pairs of mass-radius ($\bm{R}$), mass-tidal deformability ($\bm{\Lambda}$), or both ($\bm{R\Lambda}$) as input parameters. We varied the input noise to create three different datasets for each input size, as summarized in Table \ref{dataset}. These sets represent different levels of noise scatter in the input data. Consequently, a total of 27 NN models were trained, each corresponding to one of these possible combinations. We conducted a grid search to optimize the model architecture specifically for the $R_0$, $\Lambda_0$, and $R\Lambda_0$ sets. Once the optimal architecture was identified, we evaluated the model's performance using confusion matrices and key metrics such as accuracy and F1 score. To ensure that the model's performance is not biased by a fixed threshold, we also calculated the AUC for accuracy as a function of the threshold. The AUC values demonstrated consistent behavior with the accuracy measured at the standard threshold of $0.5$ . 
Our findings reveal that as input noise increases, the model requires a larger input size to maintain high accuracy. In contrast, for datasets without noise, the model quickly reaches peak accuracy with a smaller number of input pairs. This suggests that for observational data—roughly equivalent to the sets with $\mathbb{X}_2$ noise levels—larger input sizes are needed to achieve accuracy comparable to the noise-free case. Nevertheless, our model achieves high scores across all evaluated metrics, including in scenarios with the highest input noise. 
We further tested the model on a different dataset, also generated using an RMF approach but with density-dependent couplings and a different particle composition. The results demonstrated good accuracy across all models, consistently exceeding 73\%.
For the AUC value and the F1 score, the performance decreased slightly.
The F1 score is particularly significant given the highly imbalanced nature of this test set, and has indicated that the model is effectively handling and correctly predicting outcomes despite the data imbalance. While the model performs well across most datasets, there is room for improvement, particularly in handling the $\Lambda$ datasets with higher noise levels.
Finally, we tested our models on real observational data for the $\bm{R}$ models with $Q=5$, using a subset of observations that lie within the model's training region. The results revealed that the majority of predictions fell within the region without hyperons, suggesting that hyperons may be absent inside NSs. More specifically,   65\% of the data across all datasets indicated predictions in the region without hyperons. We also conducted tests with fewer samples for the $R_2$ set at $Q=10$ and $Q=15$, achieving solid results that consistently indicate the absence of hyperons.
{The interpretability of machine learning models is an area of active research. In future work, we plan to conduct feature importance analyses on our current model to identify the regions of the $M(R)$ diagram that are most informative for detecting hyperonic degrees of freedom.}

Looking ahead, upcoming observations will enable us to test models that incorporate tidal deformability, potentially providing additional insights into NS compositions. A significant direction for future work involves the application of deep ensembles or the incorporation of BNNs. These approaches would allow us to quantify both epistemic and aleatoric uncertainties, providing a more comprehensive understanding of model reliability. Given the results obtained with different nuclear models it could also be advantageous to perform a more rigorous model optimization, such as using Bayesian search methods, to enhance performance and accuracy. However, this approach can be resource-intensive in terms of both time and memory. Additionally, expanding the model to consider the presence of a quark phase in NS compositions and training and testing it under these new conditions could offer intriguing insights into the physics of dense matter.
{The codes to reproduce the present study are available on request.} 
\section*{ACKNOWLEDGMENTS} 
This work was partially supported by national funds from FCT (Fundação para a Ciência e a Tecnologia, I.P, Portugal) under the projects 2022.06460.PTDC with the  DOI identifier 10.54499/2022.06460.PTDC, and UIDB/04564/2020 and UIDP/04564/2020, with DOI identifiers 10.54499/UIDB/04564/2020 and 10.54499/UIDP/04564/2020, respectively.

\bibliographystyle{apsrev4-1}
%\bibliography{biblio}

\end{document}